\documentclass[12pt]{iopart}

\usepackage{graphicx}
\usepackage{amsfonts}

\usepackage{epsfig}
\usepackage{graphicx}
\usepackage{dcolumn}
\usepackage{bm}
\usepackage{times}
\usepackage{xcolor}

\begin{document}

\title{Suppressing epidemic spreading in multiplex networks with social-support}

\author{Xiaolong Chen$^{1,2}$, Ruijie Wang$^{1,3}$, Ming Tang$^{1,2,4}$, Shi-Min Cai$^{1,2}$,
H. Eugene Stanley$^{5}$,Lidia A. Braunstein$^{5,6}$}

\address{$^{1}$ Web Sciences Center, University of Electronic
Science and Technology of China, Chengdu 611731, China}

\address{$^{2}$ Big data research center, University of Electronic
Science and Technology of China, Chengdu 610054, China}

\address{$^{3}$ A Ba Teachers University, A Ba 623002, China}

\address{$^{4}$ School of Information Science Technology, East China Normal
  University, Shanghai 200241, China}

\address{$^{5}$ Center for Polymer Studies and Department of Physics,
  Boston University, Boston, Massachusetts 02215, USA}

\address{$^{6}$ Instituto de Investigaciones F\'{i}sicas de Mar del Plata
  (IFIMAR)-Departamento de F\'{i}sica, Facultad de Ciencias Exactas y
  Naturales, Universidad Nacional de Mar del Plata-CONICET, Funes 3350,
  (7600) Mar del Plata, Argentina.}

\address{E-mail: tangminghuang521@hotmail.com}

\begin{abstract}
Although suppressing the spread of a disease is usually achieved by
investing in public resources, in the real world only a small percentage
of the population have access to government assistance when there is an
outbreak, and most must rely on resources from family or friends. We
study the dynamics of disease spreading in social-contact multiplex
networks when the recovery of infected nodes depends on resources from
healthy neighbors in the social layer. We investigate how degree
heterogeneity affects the spreading dynamics. Using theoretical analysis
and simulations we find that degree heterogeneity promotes disease
spreading. The phase transition of the infected density is hybrid and
increases smoothly from zero to a finite small value at the first
invasion threshold and then suddenly jumps at the second invasion
threshold. We also find a hysteresis loop in the transition of the
infected density. We further investigate how an overlap in the edges
between two layers affects the spreading dynamics. We find that when the
amount of overlap is smaller than a critical value the phase transition
is hybrid and there is a hysteresis loop, otherwise the phase transition
is continuous and the hysteresis loop vanishes. In addition, the edge
overlap allows an epidemic outbreak when the transmission rate is below
the first invasion threshold, but suppresses any explosive transition
when the transmission rate is above the first invasion threshold.
\end{abstract}



 \maketitle
\tableofcontents
\section{Introduction} \label{sec:intro}

An outbreak of such diseases as SARS \cite{meyers2005network} and H5N1
\cite{yuen1998clinical,de2006fatal} puts at risk the lives of countless
people. During the first nine months of the recent Ebola epidemic there
were 4507 confirmed or probable cases of infection and 2296 deaths
\cite{team2014ebola}. Increasing the investment of public resources to
control a disease pandemic can be a serious economic burden, especially
in developing countries
\cite{gallup2001economic,kirigia2009economic}. Many researches have been
done on how to optimize scarce public health care and immunization
resources when attempting to control an epidemic
\cite{stinnett1996mathematical,wang1999role,zaric2002dynamic,brandeau2005allocating},
the goal being to minimize the number of infected individuals by
determining that optimal allocation \cite{brandeau2003resource}.

A complex network science approach is now being widely used to determine
the impact of resource investment on spreading dynamics. B{\"o}ttcher
\emph{et al.} \cite{bottcher2015disease} studied the impact of resource
constraints on epidemic outbreaks and found that when the resources
generated by the healthy population cannot cover the costs of healing
the infected population the epidemics go out of control and
discontinuous transitions
\cite{araujo2010explosive,nagler2012continuous,d2015anomalous,
  chen2016crossover,boccaletti2016explosive} occur. Chen \emph{et al.}
\cite{chen2016critical} explored the critical influence of resource
expenditure on constraining epidemic spreading in networks and found
that public resources can affect the stability of the disease
outbreak. At a certain disease transmission rate there is a critical
resource level above which a discontinuous phase transition in the
infected population occurs. B{\"o}ttcher \emph{et al.}
\cite{bottcher2016connectivity} assumed that only the
central nodes in a network can provide the necessary
care resource, and they found that a discontinuous transition in
infected nodes occurs when the central nodes are
surrounded by infected nodes. All of these researches focus on
how public resource investment affects the spread of disease.

In real-world scenarios only a small percentage of patients are assisted
by public resources. The majority depend on help from family and friends
who provide economic
\cite{seeman1996social,schulz2008physical,drummond2015methods} and
emotional support \cite{cohen1985social,thoits1995stress}. We thus study
how social support from family and friends affects the dynamics of
disease spreading. In a social network, a node has different
connections in different settings. We can thus regard friendship ties
(virtual contacts) and co-worker ties (physical contacts) as two
different network layers.  Although economic and medical resources and
sources of information usually propagate through social relationships,
diseases usually propagate through physical contacts. Thus we use a
multiplex network of two-layers
\cite{gomez2013diffusion,granell2013dynamical,bianconi2016percolation,de2016physics}
to study how resource allocation in the social layer affects the
spreading dynamics in the contact layer.

We use the susceptible-infected-susceptible (SIS) model in a
multiplex network of two-layers to mimic the coupling dynamics between disease
spreading and resource support. The disease propagates through the layer
of physical contacts, but infected nodes seek help from their neighbors
through the layer of social relations. Infected nodes receive resources
from healthy neighbors and do not generate resources. We analyze the
process using a dynamic message passing (DMP) approach
\cite{karrer2010message,shrestha2014message,shrestha2015message,wang2016unification}.
We examine how degree heterogeneity affects the dynamical process and
find that the infected density in the steady state ($\rho$) increases
continuously at the first epidemic threshold and then jumps suddenly at
the second threshold. Hysteresis loops exist in the phase transition of
the infected density, and the size of the hysteresis region and the value of the
invasion threshold decrease with the degree heterogeneity. Examining
how edge overlap between the two layers affects the dynamics of
spreading we find that the overlap has a critical
value. When the overlap is below the critical value,
the infected density first increases continuously
and then discontinuously with disease transmission rate,
and there are hysteresis loops. When the overlap is above
the critical value, the phase transition of $\rho$ is continuous and
there is no hysteresis loop. We also find that when the transmission
rate is below the first invasion threshold the disease
outbreaks more easily for a large edge overlap,
but when the transmission rate is above the first invasion threshold the
edge overlap suppresses the disease spreading and the second invasion
threshold increases as the overlap increases.

\section{Epidemic model with social-support} \label{sec:model}

\begin{figure}
\begin{center}
\includegraphics[width=0.55\linewidth]{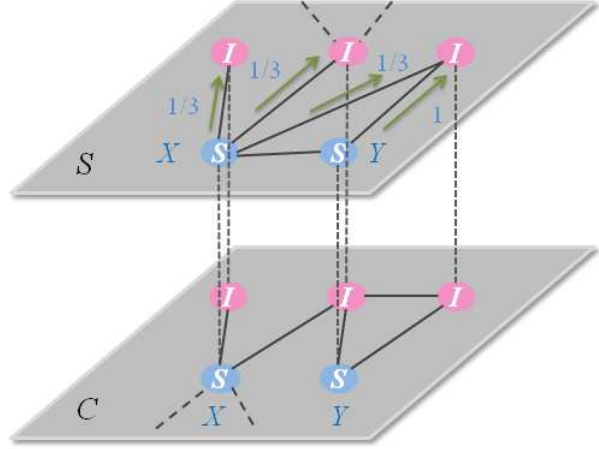}
\caption{(Color online) Schematic diagram of resource
    allocation in a multiplex network. The upper layer represents the
  social relationship network where the healthy individuals (purple
  nodes) would equally allocate their personal resource to the immediate
  infected neighbors (red nodes), as denoted by the arrows. The lower
  layer represents the physical contact network where the epidemic
  spreading takes place.}
\label{Schematic}
\end{center}
\end{figure}

In a multiplex network of two-layers, each layer has $N$ nodes
and each node in the first layer has a counterpart
in the second layer. Here the upper
layer is the social relationship network (e.g., Facebook friends and
family members) from which healthy nodes allocate resources to infected
neighbors (see layer $S$ in Fig.~\ref{Schematic}). The lower layer is
the physical contact network through which the disease spreads (see
layer $C$ in Fig.~\ref{Schematic}). Variables $A$ and $B$ are the
adjacency matrices of layer $S$ and $C$ with elements $a_{ij}$ and
$b_{ij}$. If nodes $i$ and $j$ are connected by one edge in layer $S$,
$a_{ij}=1$, otherwise $a_{ij}=0$. The same is true in layer
$C$. We denote by $s_\upsilon$ the node state variable of
  node $\upsilon$, and if it is in the susceptible state $s_\upsilon=0$,
  otherwise $s_\upsilon=1$. We assume that each healthy individual has
a certain resource level $r$ per unit time, which for simplicity we set
at $r=1$. Resources are distributed equally to infected neighbors.
Figure~\ref{Schematic} shows that node $X$ distributes one resource unit
to three infected neighbors in layer $S$, and that node $Y$ distributes
one resource unit to one infected neighbor.
For the sake of analytical tractability, we assume that the total
resource is not cumulative in the system, and if healthy
nodes do not allocate their resources to neighbors they consume these
resources themselves. In addition, infected nodes consume all of the received
resources at the current time step, and each healthy individual
generates a new one-unit resource at the next time step.
Using this definition, the resources that node $j$ gives to
node $i$ in layer $S$ is
\begin{equation}
R_{j\rightarrow i}=\frac{1}{\sum_{\upsilon} a_{j\upsilon}s_{\upsilon}}.
\label{resAlloc}
\end{equation}
Without resource support a node recovers spontaneously at a rate $\mu_0$
\cite{valdez2016failure}, and for simplicity we assume $\mu_0=0$.  The
recovery rate of $i$ at time $t$ is
\begin{equation}
\mu_i(t)=\mu_r\frac{R_{i}(t)}{k_{i}^{S}},
\label{recovery}
\end{equation}
where $\mu_i(t)\equiv\mu(R_i(t))$, and $R_i(t)$ is the
expected resources that node $i$ receives from healthy neighbors. The
$\mu_r$ value is the coefficient that represents the efficiency of
resource support from neighbors, $\mu_r\in[0,1]$, and $k_{i}^{S}$ is the
degree of $i$ in layer $S$.
The recovery rate of infected nodes is assumed to
be positively related to the resource
received from healthy neighbors in layer $S$. In
real-world setting the cost of repairing a vital node in a complex
system is much higher than the cost of repairing a common
node. For example, because hub airports in airline networks play a vital
role in connecting a large number of countries and regions, the
repairing cost when they fail is much higher than that for
lower-degree airports \cite{guimera2005worldwide}.
Similarly, the cost of repairing hub nodes in brain networks is much
higher than the cost of repairing common nodes
\cite{bullmore2012economy}. The same is true in epidemic
spreading. Individuals exposed to viruses over a long period of time,
e.g., medical staff members who are in constant contact with infected
individuals, have large degrees in physical contact networks. Community
leaders are also hub nodes in high-degree physical contact networks. In
both cases the cost of curing these hub nodes being infected
is much higher than other infected nodes in the contact
networks. Thus we assume that
the recovery rate of an infected node is negatively related to its degree.

We use the classical SIS model to investigate the spreading process in
multiplex networks. Each individual can be either infected or
susceptible.  Susceptible individuals are healthy and are then infected
by an infected neighbor at a rate $\beta$. Infected individuals recover
at a rate $\mu_i(t)$, which is assumed to be independent
of the availability of social resources in previous researches
\cite{PastorSatorras2001,pastor2015epidemic}.

\section{Dynamic message-passing method}

We use dynamic message-passing method to analyze the spreading dynamics. In
this method a variable ``message'' passes through the directed edges of
the network and does not backtrack to the source node. Our message is
$\theta_{j\rightarrow i}$, the probability that node $j$ is infected by
its neighbors other than $i$. In addition, $\rho_i(t)$ is the
probability that node $i$ is in the infected state at time $t$. The
probability that an infected node $i$ will connect to a healthy node $j$
in layer $S$ is $a_{ij}(1-\theta_{j\rightarrow i}(t))$, and the expected
number of infected neighbors of node $j$ is $\sum_{\ell\neq
  i}a_{j\ell}\theta_{\ell\rightarrow j}(t)+1$, where the plus one takes
 into account that node $i$ is infected. Thus the resource
$R_i(t)$ that node $i$ receives from healthy neighbors is
\begin{equation}
R_i(t)=\sum_{j}a_{ij}(1-\theta_{j\rightarrow
  i}(t))\frac{1}{\sum_{\ell\neq i}a_{j\ell}\theta_{\ell\rightarrow
    j}(t)+1}.
\label{resource}
\end{equation}
Using this definition, the discrete-time version of evolution of
$\rho_i(t)$ \cite{gomez2010discrete} is
\begin{equation}
\rho_i(t+\Delta t)=(1-\rho_i(t))(1-q_i(t))+(1-\mu_i(t))\rho_i(t),
\label{dynamic}
\end{equation}
where $\Delta t$ is the time increment, which we set at
  $\Delta t=1$, and $q_i(t)$ is the probability that $i$ is not infected by
any neighbor in layer $C$, which is given by
\begin{equation}\label{c}
q_i(t)= \prod_{j\in \mathcal{N}^{C}_{i}}(1-\beta\theta_{j\rightarrow i}(t)),
\end{equation}
where $\mathcal{N}^{C}_{i}$ is the neighbor set of $i$ in layer
$C$. Note that to exclude any contribution of node $i$ to the infection
of $j$, we adopt $\theta_{j\rightarrow i}(t)$ instead of $\rho_j(t)$ in
Eq.~(\ref{c}). Similarly, the discrete-time version of evolution of
$\theta_{j\rightarrow i}(t)$ is
\begin{equation}
\theta_{j\rightarrow i}(t+1)=(1-\theta_{j\rightarrow i}(t))(1-\phi_{j\rightarrow i}(t))+(1-\mu_j(t))\theta_{j\rightarrow i}(t).
\label{cavityRate}
\end{equation}
Here $(1-\phi_{j\rightarrow i}(t))$ is the probability that $j$ is
infected by at least one neighbor other than $i$. Thus
$\phi_{j\rightarrow i}(t)$ is
\begin{equation}
\phi_{j\rightarrow i}(t)= \prod_{\ell\in \mathcal{N}^{C}_{j}\setminus
  i}(1-\beta\theta_{\ell\rightarrow j}(t)).
\label{seccavity}
\end{equation}
Here $\mathcal{N}^{C}_{j}\setminus i$ is the neighbor set of $j$
excluding $i$, and the fraction of infected nodes at time $t$ is
\begin{equation}
\rho(t)=\frac{1}{N}\sum^{N}_{i=1}\rho_i(t),
\label{density}
\end{equation}
where $\rho_i(\infty)\equiv \rho_i$ and $\mu_i(t)\equiv\mu_i$
at the steady state $t\rightarrow\infty$.
Solving Eqs.~(\ref{dynamic}) and
(\ref{cavityRate}) at the stationary state
\begin{equation}\label{steadyRho}
  \rho_i=(1-\rho_i)(1-q_i)+(1-\mu_i)\rho_i
\end{equation}
and
\begin{equation}
  \theta_{j\rightarrow i}=(1-\theta_{j\rightarrow
    i})(1-\phi_{j\rightarrow i})+(1-\mu_j)\theta_{j\rightarrow i},
  \label{steadyTheta}
\end{equation}
we obtain the phase diagram of the model. We use iteration to
numerically compute the evolution of the state of network nodes.

Due to nonlinearities in Eqs.~(\ref{resource})--(\ref{seccavity}) they
do not have a closed analytic form, and this disallows obtaining the
epidemic threshold $\beta_c$. If $\beta>\beta_c$, $\rho>0$, otherwise
$\rho=0$ in the steady state. When $\beta\rightarrow\beta_c$,
$\rho_i\rightarrow0$, $\theta_{j\rightarrow i}\rightarrow0$, and the
number of infected neighbors of each healthy node in layer $S$ is
approximately zero in the thermodynamic limit, prior to reaching the
epidemic threshold $(1-\theta_{j\rightarrow i})\rightarrow1$. If we add
these assumptions to Eq.~(\ref{resource}) resource $R_i$ becomes
$R_i\rightarrow k^S_i$, we will obtain the recovery rate
$\mu_i\rightarrow \mu_r$ in the steady state [see
Figs.~\ref{Evolution(struct)}(a) and \ref{Evolution(edgelap)}(a)].

To compute the threshold, we linearize Eqs.~(\ref{cavityRate}) and
(\ref{seccavity}) around $\theta_{j\rightarrow i}=0$ and obtain
\begin{equation}\label{cc}
  q_i\approx1-\beta\sum_{j=1}^{N}{b_{ji}\theta_{j\rightarrow
      i}},
\end{equation}
and
\begin{equation}
  \phi_{j\rightarrow i}\approx1-\beta\sum{\mathbf{B}_{j\rightarrow
      i,l\rightarrow h}\theta_{l\rightarrow h}},
\label{rseccavity}
\end{equation}
where $\mathbf{B}$ is the non-backtracking matrix
\cite{krzakala2013spectral} of layer $C$ and
\begin{equation}\label{nonbacking}
\mathbf{B}_{j\rightarrow i,l\rightarrow h}=\delta_{jh}(1-\delta_{il}),
\end{equation}
where $\delta_{il}$ is a Dirac delta function. Inserting
Eq.~(\ref{rseccavity}) into Eq.~(\ref{steadyTheta}) and neglecting
second-order terms we obtain
%
\begin{equation}
 \sum{(-\delta_{lj}\delta_{ih}\mu_r+\beta\mathbf{B}_{j\rightarrow
    i,l\rightarrow h})}\theta_{l\rightarrow h}=0.
 \label{rcavityRate}
\end{equation}
To solve Eq.~(\ref{rcavityRate}) we define a $2E\times2E$
matrix $\mathbf{J}$, where $E$ is the number of edges and the elements
of $\mathbf{J}$ are
\begin{equation}\label{jacobian}
  \mathbf{J}_{j\rightarrow i,l\rightarrow
    h}=-\delta_{lj}\delta_{ih}\mu_r+\beta\mathbf{B}_{j\rightarrow
    i,l\rightarrow h}.
\end{equation}

The system enters a global epidemic region in which the epidemic grows
exponentially when the largest eigenvalue of $\mathbf{J}$ is greater
than zero
\cite{shrestha2015message,pastor2015epidemic,wang2016unification}. Thus
we can obtain the epidemic threshold as
\begin{center}
\begin{equation}
  \beta_c=\frac{1}{\Lambda_J},
\label{threshold}
\end{equation}
\end{center}
where $\Lambda_J$ is the largest eigenvalue of $\mathbf{J}$.

\section{Numerical and simulation results}

To examine how resource support affects epidemic dynamics, we perform
numerical computations and stochastic simulations in the networks.
Because many real-world complex networks have a highly skewed
degree distribution, e.g., Facebook
\cite{viswanath2009evolution} and the World Wide Web
\cite{adamic2000power}, we focus on networks with a heterogenous degree
distribution. We assume that the two layers of the network have the same
degree sequences ($k_i^{S}=k_i^{C}$). Thus for simplicity we denote
$k_i$ to be the degree of node $i$ in both layers $S$ and $C$.

To build our multiplex network we use an uncorrelated configuration
model (UCM) \cite{catanzaro2005generation} with a given degree
distribution $P(k)\sim k^{-\gamma}$ in which $\gamma$ is the degree
exponent. Here a smaller $\gamma$ implies a more heterogeneous degree
distribution.  The maximum degree is determined by the structural
cut-off $k_{\rm max}\sim \sqrt{N}$ \cite{boguna2004cut} and we set the
minimum degree at $k_{\rm min}=3$. In addition we disallow
multiple and self-connections and set the network size as
$N=10000$. When studying resource support from neighbors, we eliminate
any possibility of spontaneous recovery, i.e., $\mu_0=0$, and assume
that node recovery is solely dependent on the amount of resources
received. Here we set the efficiency parameter at $\mu_r=0.6$
and the $\mu_r$ value does not affect the result
  \cite{pastor2001epidemicprl,pastor2015epidemic}.

To determine the epidemic threshold, we use a susceptibility measure
\cite{ferreira2012epidemic,shu2016recovery}
\begin{equation}
\chi = N\frac{\langle\rho^2\rangle-\langle\rho\rangle^2}{\langle\rho\rangle},
\end{equation}
where $\langle\ldots\rangle$ is the ensemble averaging, and $\chi$
exhibits peaks at the transition points.

We now examine how degree heterogeneity and edge overlap between the two
layers of the network affect its dynamic features.

\subsection{Effects of degree heterogeneity}


\begin{figure}
\begin{center}
\includegraphics[width=0.5\linewidth]{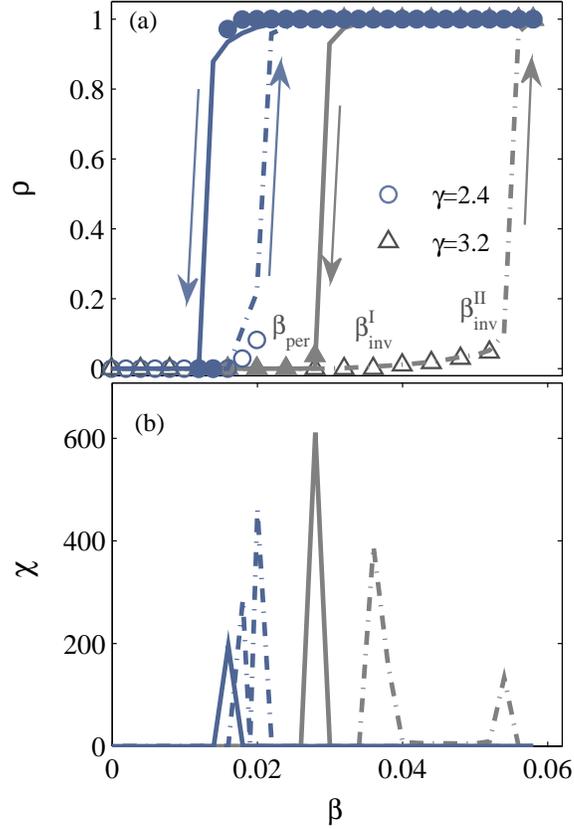}
\caption{(Color online) Influence of degree heterogeneity on the
  spreading dynamics. (a) The infected density $\rho$ vs disease
  transmission rate $\beta$ for $\gamma=2.4$ (denoted by blue circles)
  and $\gamma=3.2$ (denoted by dark grey trangles). The dotted lines and
  solid lines are analytical results and arrows indicate the direction
  of the hysteresis loop. (b) Susceptibility measure $\chi$ vs $\beta$
  for $\gamma=2.4$ and $\gamma=3.2$, the dotted lines and solid lines
  correspond to the cases for the initial infected density $\rho_0=0.01$
  and $\rho_0=0.9$. Quantities $\beta_{\rm inv}$ and $\beta_{\rm per}$
  are the invasion and persistence thresholds, and $\beta_{\rm
    inv}^{I}$, $\beta_{\rm inv}^{II}$ represent the first and second
  invasion thresholds respectively.}
\label{struct(capa)}
\end{center}
\end{figure}

To investigate how degree heterogeneity affects spreading dynamics, we
disallow any edge overlap between the two layers, i.e., nodes are
randomly connected by edges in layer $S$ and layer $C$, and the amount
of edge overlap $m_e$ is approximately 0 in the thermodynamic limit.

To examine $\rho$ as a function of $\beta$, we randomly select one
percent of the nodes to be seeds ($\rho(0)=0.01$).
Figure~\ref{struct(capa)}(a) shows the epidemic spreading for
$\gamma=2.4$ and $\gamma=3.2$. Note the hybrid phase transition in
$\rho$ that exhibits properties of both continuous and
  discontinuous phase transitions. As $\beta$ increases $\rho$ grows
continuously at $\beta_{\rm inv}^{I}$.  Then an infinitely small
increase in $\beta$ induces an sudden jump of $\rho$ at
 $\beta_{\rm inv}^{II}$, where $\beta_{\rm
  inv}^{I}$ and $\beta_{\rm inv}^{II}$ are the first and second invasion
thresholds. The $\rho$ transition type indicates that there are three
possible system states, (i) completely healthy, (ii) partially infected,
and (iii) completely infected.  This differs significantly from the
classical SIS model. In addition, we find hysteresis loops in the phase
transition of $\rho$ when $\gamma=2.4$ and $\gamma=3.2$ [see
  Fig.~\ref{struct(capa)}(a)]. When the seed density is
  initially low, e.g., $\rho(0)=0.01$, the disease breaks out at the
  invasion threshold $\beta_{\rm inv}^{I}$, but when it is initially
  high, e.g., $\rho(0)=0.9$, the disease breaks out at the persistence
  threshold $\beta_{\rm per}$.  The arrows in Fig.~\ref{struct(capa)}(a)
  indicate the direction of the hysteresis loops. We determine critical
points $\beta_{\rm inv}^{I}$ and $\beta_{\rm inv}^{II}$ and persistence
threshold $\beta_{\rm per}$ using the susceptibility $\chi$ shown in
Fig.~\ref{struct(capa)}(b). The theoretical results obtained from the
numerical iterations agree with the simulation results [see the lines in
  Fig.~\ref{struct(capa)}(a)].

\begin{figure}
\begin{center}
\includegraphics[width=0.5\linewidth]{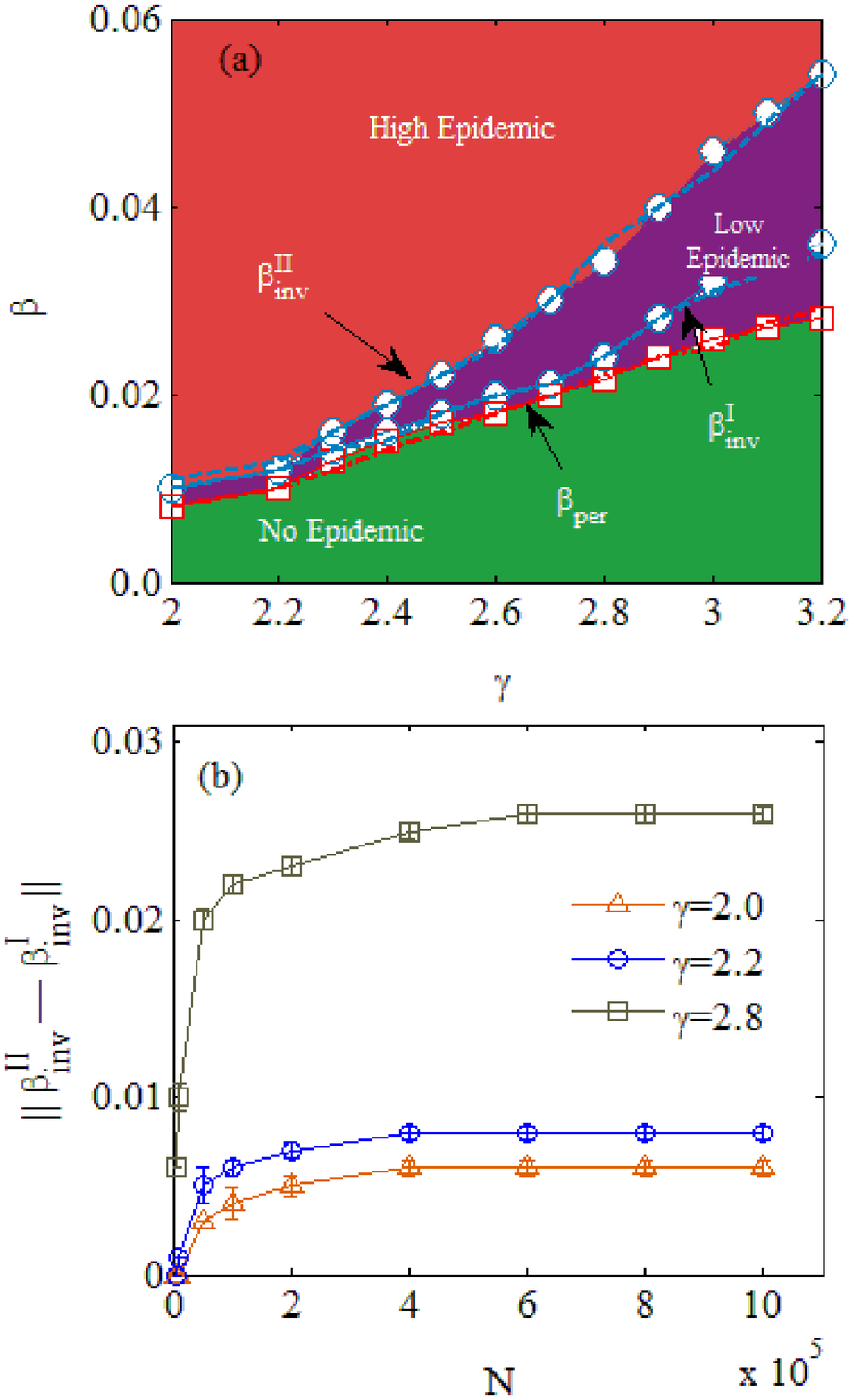}
\caption{(Color online). Effect of degree heterogeneity on spreading
  dynamics. (a) Phase diagram in the two-parameter $(\beta,\gamma)$
  space. Three regions of the stable state are obtained. The high
  epidemic region with large value of $\rho$ (denoted by red color) in
  steady state, the no epidemic region with zero value of $\rho$
  (denoted by green color), and the low epidemic region with small value
  of $\rho$ (part of the purple region bounded by the two critical
  lines). The hysteresis region (denoted by purple color) is bounded
  within $\beta_{\rm inv}^{II}$ and $\beta_{\rm per}$ (denoted by red
  squares). The two invasion thresholds $\beta_{\rm inv}^{I}$ (denoted
  by lower blue circles), $\beta_{\rm inv}^{II}$ (denoted by upper blue
  circles) and persistence thresholds $\beta_{\rm per}$ are determined
  by the susceptibility measure $\chi$. Theoretical results obtained
  from the DMP method are denoted by dotted lines in the figure. (b) The
  thresholds interval $\|\beta_{\rm inv}^{II}-\beta_{\rm inv}^{I}\|$ is
  plotted as a function of system size $N$ for three different values of
  $\gamma$: $\gamma=2.0$ (red triangles), $\gamma=2.2$ (blue circles),
  and $\gamma=2.8$ (dark grey squares). Error bars are smaller than the
  symbols used for the data points.}
\label{hysteresis(struct)}
\end{center}
\end{figure}

We next determine how degree heterogeneity (i.e., parameter $\gamma$)
influences the spreading dynamics.
Figure~\ref{hysteresis(struct)}(a) shows the two-parameter
$(\beta,\gamma)$ phase diagram.  The parameter space is partitioned into
three regions according to $\rho$ value.  When $\beta<\beta_{\rm
  inv}^I$, the system falls into the no-epidemic regime, i.e.,the green
and part of the purple area below $\beta_{\rm inv}^I$.  When $\beta_{\rm
  inv}^I\leq\beta<\beta_{\rm inv}^{II}$, it falls into the low-epidemic
regime (bounded by two critical lines) in which $\rho$ increases slowly
with $\beta$.  Finally, above $\beta_{\rm inv}^{II}$, $\rho$ suddenly jumps
to the high epidemic regime (red) in which approximately all nodes are
infected. The regime between invasion threshold $\beta_{\rm inv}^{II}$
and persistence threshold $\beta_{\rm per}$ is the hysteresis region
(purple). The values of $\beta_{\rm inv}^{I}$ and $\beta_{\rm
  inv}^{II}$ both increase with $\gamma$. Although we can
obtain the theoretical value of $\beta_{\rm inv}^{I}$ from
Eq.~(\ref{threshold}), we cannot obtain the theoretical value of
$\beta_{\rm inv}^{II}$ and $\beta_{\rm per}$ by linearizing the
equations around $\rho_i\rightarrow 0$ and $\theta_{j\rightarrow
  i}\rightarrow0$ and thus we must apply numerical methods using
Eqs.~(\ref{dynamic}) and (\ref{cavityRate}).  We first define a judgment
value $\epsilon$ that is linear with system size
$N$. Without loss of generality we set $\epsilon=0.3$.  We then define
the jump size $\Delta\rho$ to be
\begin{equation}
  \Delta\rho=\rho(\beta)-\rho(\beta-\Delta\beta),
\label{jump}
\end{equation}
where $\Delta\beta$ is an infinitesimal increment in $\beta$, which we
set at $\Delta\beta=0.001$, and $\rho(\beta)$ is the infected density in
the steady state when the transmission rate is $\beta$. We obtain the
threshold when $\Delta\rho\geq\epsilon$ is at a certain $\beta$ value in
the thermodynamic limit \cite{nagler2011impact,chen2016crossover}. Using
the numerical method, we obtain the second invasion threshold
$\beta_{\rm inv}^{II}$ and the persistence threshold $\beta_{\rm per}$.
Figure~\ref{hysteresis(struct)} shows that the theoretical values marked
by dotted lines agree with the simulation results. The change in
the system state among the three regions indicates that the phase
transitions of $\rho$ are hybrid.  Figure~\ref{hysteresis(struct)}(a)
shows that the low epidemic and hysteresis regions expand as $\gamma$
increases.

To demonstrate that there are two invasion thresholds in networks with
heterogeneous degree distribution, we use a finite-size scaling analysis
\cite{newman1999monte}.  Figure~\ref{hysteresis(struct)}(b) shows the
interval in $\beta_{\rm inv}^I\leq\beta<\beta_{\rm inv}^{II}$, which we
denote $\|\beta_{\rm inv}^{II}-\beta_{\rm inv}^{I}\|$, as a function of
$N$ for $\gamma=2.0$, $\gamma=2.2$, and $\gamma=2.8$, where
$\|\bullet\|$ is the norm operator.  Figure~\ref{hysteresis(struct)}(b)
shows the values of $\|\beta_{\rm inv}^{II}-\beta_{\rm inv}^{I}\|$
converging asymptotically to positive constant values in the
thermodynamic limit, i.e., ${\lim_{N\to \infty} \|\beta_{\rm
    inv}^{II}-\beta_{\rm inv}^{I}\|}\simeq 0.006$ for $\gamma=2.0$,
${\lim_{N \to \infty} \|\beta_{\rm inv}^{II}-\beta_{\rm
    inv}^{I}\|}\simeq 0.008$ for $\gamma=2.2$, and ${\lim_{N \to \infty}
  \|\beta_{\rm inv}^{II}-\beta_{\rm inv}^{I}\|}\simeq 0.026$ for
$\gamma=2.8$, which implies the two invasion thresholds do not merge
when $\gamma\leq 2.2$ and the two are always present in networks with a
heterogeneous degree distribution.

To analyze the sudden jump of $\rho$ and the hysteresis loops, we
examine the transmission process analytically using mean-field
approximation in random regular networks (RRNs), which corresponds to
the limit $\gamma\rightarrow\infty$. Through a
bifurcation analysis we account for the existence of the sudden jump of
$\rho$ and the hysteresis loops (see appendix information).  Note that
the first threshold $\beta_{\rm inv}^I$ disappears in the RRNs and the
transition of $\rho$ is discontinuous when it is not hybrid [see
  Fig.~\ref{RRN(capa)}(a)].

\begin{figure}
\begin{center}
\includegraphics[width=0.8\linewidth]{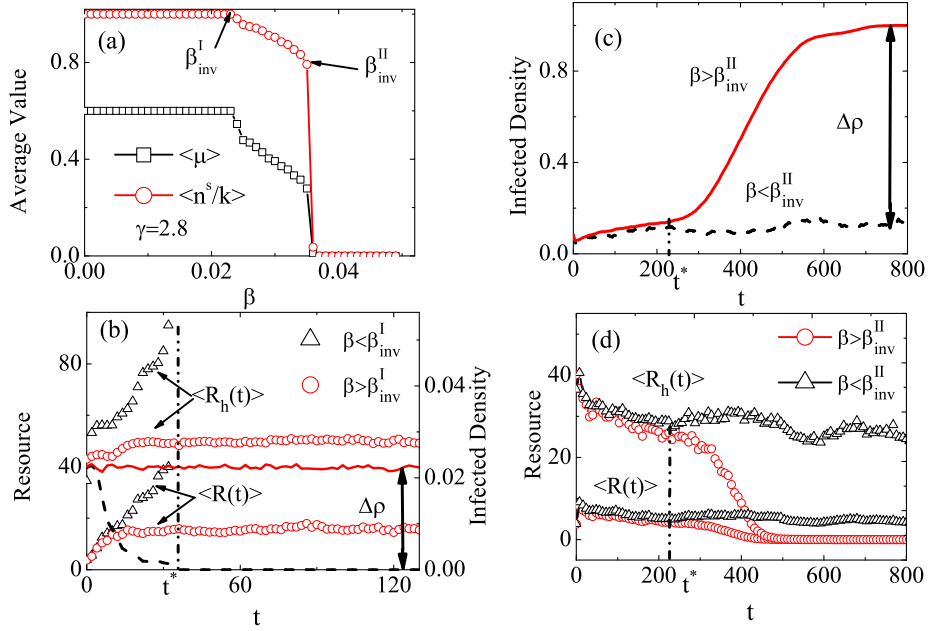}
\caption{(Color online) Analysis of the hybrid transitions in
  heterogeneous multiplex networks. (a) Plot of fraction of healthy (S)
  neighbors around infected nodes $\langle n_s/k\rangle$ (red circles)
  and average recovery rate $\langle\mu\rangle$ (black squares) as
  functions of $\beta$ in the steady state for $\gamma=2.8$, where
  initial intected density is $\rho(0)=0.01$. (b) Time evolution of
  average resource of all infected nodes $\langle R(t)\rangle$ and hub
  nodes $\langle R_h(t)\rangle$ (left ordinate) for $\beta>\beta_{\rm
    inv}^{I}\simeq0.023$ (red circles) and $\beta<\beta_{\rm inv}^{I}$
  (black triangles), and the time evolutions of infected density
  $\rho(t)$ (left ordinate) for $\beta<\beta_{\rm inv}^{I}$ (lower black
  line) and $\beta>\beta_{\rm inv}^{I}$ (upper red line). $\Delta\rho$
  is the jump of $\rho$ in the steady state for $\beta$ is just below
  $\beta_{\rm inv}^{II}$ and just above $\beta_{\rm inv}^{II}$. $t^{*}$
  is the moment when all the neighbors of the infected nodes are in
  healthy state. (c) The time evolution of infected density $\rho(t)$
  for $\beta$ close to $\beta_{\rm inv}^{II}\simeq0.033$.(d) The average
  resource of $\langle R(t)\rangle$ and $\langle R_{h}(t)\rangle$ as
  functions of $t$, $t^{*}$ is the critical time when the received
  resource of the hub nodes drops abruptly, which corresponds to the
  moment when $\rho(t)$ increases sharply in (c).}
\label{Evolution(struct)}
\end{center}
\end{figure}

To explain the hybrid transition when $\gamma$ is finite, i.e., when
$\gamma\leq 3.2$, we investigate the number of susceptible neighbors around
each infected node in layer $S$ and their recovery rates as a function
of $\beta$. In the steady state the number of each infected node's
susceptible neighbors in layer $S$ is $n_{s_i}$ and their fraction
$n_{s_i}/k_i$.  Here the recovery rate is $\mu_i$. To evaluate the
collective state, we examine the average quantity $\langle n_s/k
\rangle$ of $n_{s_i}/k_i$ and the average quantity $\langle\mu\rangle$
of the recovery rate. Figure~\ref{Evolution(struct)}(a) shows plots of
$\langle n_s/k \rangle$ and $\langle\mu\rangle$ as functions of $\beta$
for $\gamma=2.8$. We find that both $\langle n_s/k \rangle$ and
$\langle\mu\rangle$ are constant when $\beta<\beta_{\rm inv}^I$, which
implies zero values for $\rho$. They then slowly decrease until they
reach the $\beta_{\rm inv}^{II}$, at which point an infinitesimal
increase in $\beta$ causes a jump in $\langle n_s/k \rangle$ and
$\langle\mu\rangle$. Figures~\ref{Evolution(struct)}(b)--\ref{Evolution(struct)}(d)
show the time dependence near $\beta_{\rm inv}^I$ and $\beta_{\rm
  inv}^{II}$. Figure~\ref{Evolution(struct)}(b) shows the time
evolution of the infected density $\rho(t)$ around $\beta_{\rm
  inv}^{I}\simeq0.023$ for $\gamma=2.8$. The difference in
$\rho(\infty)$ for $\beta$ just below and above threshold $\beta_{\rm
  inv}^{I}$ is $\Delta\rho$ [see Eq.~(\ref{jump})]. Note that
$\rho(\infty)$ increases slowly at $\beta_{\rm inv}^{I}$, i.e., a small
increment $\Delta\rho\simeq 0.022$. We next examine the time evolutions
of the average resources of the infected nodes $\langle R(t)\rangle$ and
the hub nodes $\langle R_{h}(t)\rangle$. Note that without loss of
generality we can assign hub node status to nodes with a degree larger
than $k=30$. Note also that when $\beta$ is just below $\beta_{\rm
  inv}^{I}$ both $\langle R(t)\rangle$ and $\langle R_{h}(t)\rangle$
increase until $t=t^*$, which implies that all infected nodes have
acquired sufficient resources to recover and $\rho(t)$ drops to zero. In
contrast, when $\beta$ is just above $\beta_{\rm inv}^{I}$ and the
transmission rate is low, the promotion effect of the hub nodes allows
the disease to spread on a finite scale and infected nodes have
sufficient resources for recovery. Thus infection and recovery processes are
balanced, and the values of $\langle R(t)\rangle$ and $\langle
R_{h}(t)\rangle$ fluctuate around a finite value when $t\rightarrow
t_{\infty}$ [see Fig.~\ref{Evolution(struct)}(b)]. As $\beta$ smoothly
increases at $\beta=\beta_{\rm inv}^{I}$, the level of available resources
decreases continuously as the number of infected nodes increases [see
  Figs.~\ref{Evolution(struct)}(a)]. Thus the density of infection
increases continuously at $\beta_{\rm inv}^{I}$.
Figures~\ref{Evolution(struct)}(c) and \ref{Evolution(struct)}(d) show a
critical time $t^*\simeq 220$ at which $\beta$ is approximately
$\beta_{\rm inv}^{II}\simeq0.033$. At the early stage of the propagation
process, i.e., when $t<t^*$, the disease spreads through the local seed
nodes.  Because most of neighbors of the infected nodes in layer $S$
remain healthy, they have a sufficient resource level to recover. Here
the infection and recovery processes are balanced. As the $\rho(t)$
value increases slowly the available resources levels $\langle
R_h(t)\rangle $ and $\langle R(t)\rangle$ for
$\beta\simeq\beta^{II}_{\rm inv}$ slowly decrease [see
  Figs.~\ref{Evolution(struct)}(c) and \ref{Evolution(struct)}(d)].
When $\beta<\beta^{II}_{\rm inv}$ the infection and recovery processes
remain balanced when $t\rightarrow\infty$, thus the density of infection
fluctuates around a small finite value when $t\rightarrow t_{\infty}$
($\rho(\infty)\simeq0.18$) [see Fig.~\ref{Evolution(struct)}(c)].
Because infection and recovery processes are balanced in
$\beta^{I}_{\rm inv}\leq\beta<\beta^{II}_{\rm inv}$, the value of $\rho$
increases slowly. Note that as hub nodes disappear in the RRNs the
disease is suppressed until $\beta$ reaches a threshold at which point
it jumps discontinuously, the balance disappears (see Appendix), and
only one threshold remains. When $\beta>\beta^{II}_{\rm inv}$ the
transmission rate is relatively large and the balance between infection
and recovery is broken. Infecting the healthy nodes in layer $C$
decreases the resources available to the nodes in layer $S$ and delays
the recovery of infected nodes. This recovery delay increases the
effective transmission probability in layer $C$, more healthy nodes are
infected, and both the available resources and the recovery rate
decrease. This causes a cascading infection in system nodes that is
accelerated when hub nodes are surrounded by infected nodes, and this
can cause total system failure. Figure~\ref{Evolution(struct)}(d) shows
an abrupt drop of $\langle R_h(t)\rangle$ and $\langle R(t)\rangle$ at
$t^*$ when $\beta>\beta^{II}_{\rm
  inv}$. Figure~\ref{Evolution(struct)}(c) shows a rapid increase in the
density of infection from a small value $\rho(t^*)\simeq 0.18$ to a high
value $\rho(\infty)\simeq 1.0$. Note that in the steady state the large
difference $\Delta\rho$ between $\beta<\beta_{\rm inv}^{II}$ and
$\beta>\beta_{\rm inv}^{II}$ causes explosive transitions. This explains
the hybrid transition in networks with a heterogeneous degree
distribution.

Figure~\ref{Evolution(struct)}(d) shows the evolution of the resource
level in the hub nodes. This explains the decrease in the two invasion
thresholds and the gap that appears between the two thresholds with the
increase of degree heterogeneity. A more heterogeneous network has more
hub nodes and is more sensitive to increases in $\beta$. Thus increasing
the degree heterogeneity reduces the gap between the two thresholds [see
  Fig.~\ref{hysteresis(struct)}].

These numerical and simulation results differ greatly from the classical
SIS model. In the multiplex networks with a heterogeneous degree distribution, degree
heterogeneity enhances disease spreading and the phase transition is
hybrid. Besides, there are hysteresis loops in the phase transition of $\rho$,
and the interval between the two invasion thresholds and the hysteresis
region decreases as degree heterogeneity increases. When
$\gamma\rightarrow\infty$ the network is approximately a RRN,
$\beta_{\rm inv}^I$ disappears as hub nodes disappear, and the transition is
discontinuous.

\subsection{Effects of edge overlap}

In social networks two individuals can be friends in the
social relation layer and coworkers in the physical contact layer.  In
transportation networks two cities can be connected by both an
expressway and a railway.  Thus edge overlap is essential in the
science of complex networks, especially when studying percolation
in multiplex networks \cite{cellai2013percolation}. Here we examine
how the amount of edge overlap $m_e$ between the two layers affects the
spreading dynamics.  To eliminate the effect of structure, we fix the
values $\gamma=2.2$ and $<k>=9$. We then use UCM to build a multiplex
network with two identical layers $m_e=1.0$. To generate a variety of
$m_e$ values, with a probability $q=1-m_e$ we rewire pairs
of links in layer $S$.

\begin{figure}
\begin{center}
\includegraphics[width=0.5\linewidth]{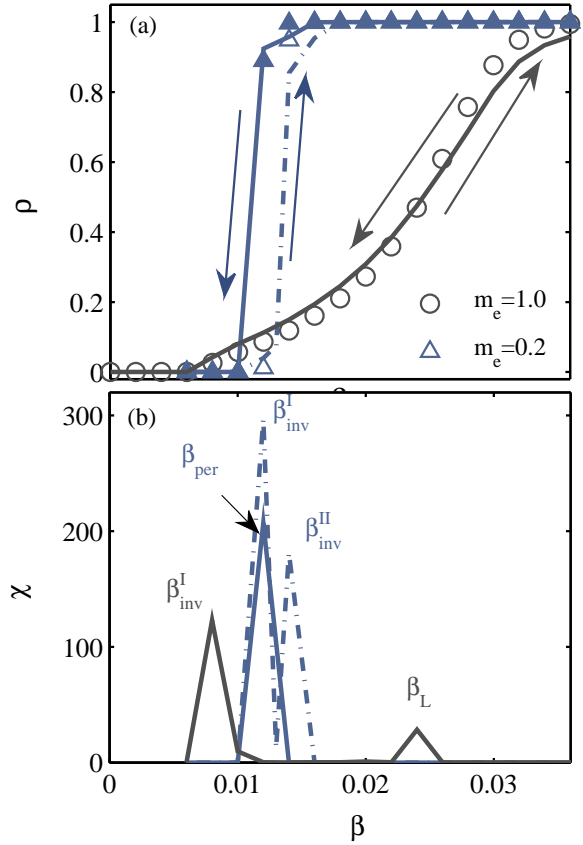}
\caption{(Color online) Influence of edge overlap on the spreading
  dynamics. (a) $\rho$ vs $\beta$ and hysteresis study for
  $m_e=1.0$ (denoted by dark grey circles), $m_e=0.2$ (denoted by blue
  trangles). The dotted lines and solid lines are analytical results,
  arrows indicate the direction of the hysteresis loop. (b)
  Susceptibility measure $\chi$ vs $\beta$ for $m_e=1.0$ and $m_e=0.2$.
  Quantities $\beta_{\rm inv}$ and $\beta_{\rm per}$ are the invasion
  and persistence thresholds, and $\beta_{\rm inv}^{I}$, $\beta_{\rm
  inv}^{II}$ respectively represent the first and second invasion thresholds,
  $\beta_{L}$ for $m_e=1.0$ represent the inflection
  point.}
\label{edgelap(capa)}
\end{center}
\end{figure}

Figure~\ref{edgelap(capa)}(a) shows a plot of $\rho$ as a function of
$\beta$ with two typical values $m_e=1.0$ and $m_e=0.2$. Note that when
the edges between the two layers overlap completely ($m_e=1.0$) the
infected density $\rho$ smoothly increases from 0 to 1 and there is no
hysteresis loop. When the rate of edge overlap between two layers is
lowered, i.e., when $m_e=0.2$, a hybrid phase transition appears.  The
infected density $\rho$ smoothly increases at $\beta=\beta_{\rm
  inv}^{I}$ and then the system acquires a low epidemic region
($\beta_{\rm inv}^{I}<\beta<\beta_{\rm inv}^{II}$) in which $\rho$
slowly increases. Subsequently at $\beta=\beta_{\rm inv}^{II}$ an
infinitesimally small increase in $\beta$ causes an abrupt jump in
$\rho$ and the disease suddenly spreads throughout the
entire system. Hysteresis loops appear in the transition process and the
arrows indicate their direction.
Figure~\ref{edgelap(capa)}(b) shows that the invasion thresholds
(i.e., $\beta_{\rm inv}^{I}$) and $\beta_{\rm inv}^{II}$ and the persistence
threshold $\beta_{\rm per}$ are determined by the susceptibility
$\chi$. Note that the hysteresis loop disappears when
$m_e=1.0$, and it no longer satisfies the definition of $\epsilon>0.3$
at $\beta_L$. Thus $\beta_L$ is an inflection point at which the
increase in $\rho$ accelerates. The theoretical results from the DMP
method agree with the simulation results.

\begin{figure}
\begin{center}
\includegraphics[width=0.5\linewidth]{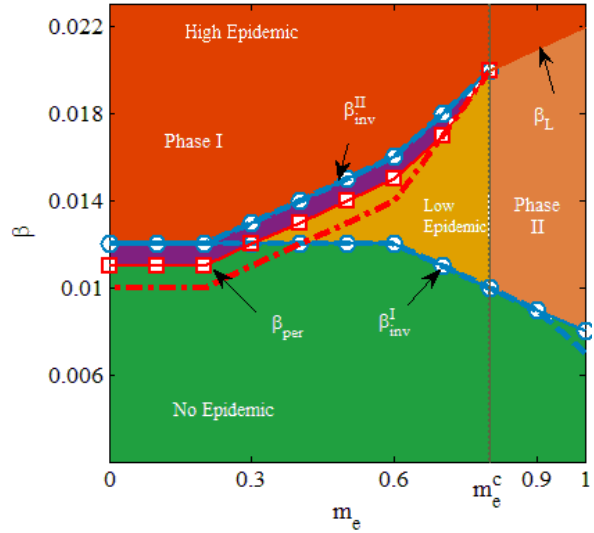}
\caption{(Color online) Phase diagram in the $(m_e,\beta)$ space.  The
  space is separated into two phase regions by the critical values
  $m_e^{II}\simeq0.8$: phase I and II. In phase region three stable
  regions are obtained: the high epidemic region denoted by red color,
  the no epidemic region denoted by green color, and the low epidemic
  region denoted by yellow color. The hysteresis region (denoted by
  purple color) is bounded with the line of $\beta_{\rm inv}^{II}$ and
  the line of $\beta_{\rm per}$ (denoted by red squares).  The invasion
  thresholds $\beta_{\rm inv}^{I}$, $\beta_{\rm inv}^{II}$ (denoted by
  blue circles) and the inflection point $\beta_L$, the persistence
  thresholds $\beta_{\rm per}$ are determined by the susceptibility
  measure $\chi$. While in phase region
  $\uppercase\expandafter{\romannumeral2}$, the phase transition of
  $\rho$ becomes continuous. The green region represents no epidemic and
  the pink region represents disease outbreaks. Theoretical results
  obtained from the DMP method are denoted by dotted lines.}
\label{hysteresis(edgeLap)}
\end{center}
\end{figure}

To determine how the amount of edge overlap between the two layers
affects the spreading dynamics, we perform simulations for values of
$m_e$ from 0 to 1 and obtain the space in the plane $(\beta,m_e)$ shown
in Fig.~\ref{hysteresis(edgeLap)}. The parameter space is separated into
phase regions I and II by a critical value of edge overlap
$m_e^{c}\simeq 0.8$.  When $m_e< m_e^{c}$ the system falls into phase I
in which the phase transition of $\rho$ is hybrid and the space is again
separated into three regions by two invasion thresholds $\beta_{\rm
  inv}^{I}$ (lower blue circles) and $\beta_{\rm inv}^{II}$ (upper blue
circles). When $\beta<\beta_{\rm inv}^{I}$ the system has a no-epidemic
region (green) in which all nodes are healthy and in a steady
state. When $\beta_{\rm inv}^{I}\leq\beta<\beta_{\rm inv}^{II}$ the
system has a low-epidemic region (orange) in which the infected density
$\rho$ increases continuously from 0 to a finite value until it reaches
the second invasion threshold $\beta_{\rm inv}^{II}$.
Figure~\ref{hysteresis(struct)}(b) shows a small low-epidemic region
when $m_e\leq0.2$ such that when $\gamma=2.2$ and $m_e=0.0$ the value of
$\|\beta_{\rm inv}^{II}-\beta_{\rm inv}^{I}\|$ converges to a non-zero
constant value when $N\rightarrow \infty$. When $\beta\geq\beta_{\rm
  inv}^{II}$ the system jumps abruptly to a high epidemic region (red)
in which the disease spreads throughout the entire
system. The hysteresis loops (purple) appear in phase I.  In
contrast, when $m_e\geq m_e^{c}$ the system falls into phase II in which
the phase transition of $\rho$ is continuous. The value of $\rho$
smoothly increases from 0 to 1 and the hysteresis loops
disappear. Figure~\ref{hysteresis(edgeLap)} shows that when
$\beta<\beta_{\rm inv}^{I}$ the value of $\beta_{\rm inv}^{I}$ decreases
as the amount of edge overlap increases. Here edge overlap promotes
disease spreading.  When $\beta\geq\beta_{\rm inv}^{I}$ the value of
$\beta_{\rm inv}^{II}$ increases as the amount of edge overlap
increases. Here edge overlap suppresses disease spreading. We obtain the
theoretical value of $\beta_{\rm inv}^{I}$ using Eq.~(\ref{threshold})
and $\beta_{\rm inv}^{II}$ and $\beta_{\rm per}$ using the method in
Section IV.A. Figure~\ref{hysteresis(edgeLap)} shows that the
theoretical values marked by the dotted lines agree with simulation
results.

\begin{figure}
\begin{center}
\centering\includegraphics[width=0.8\linewidth]{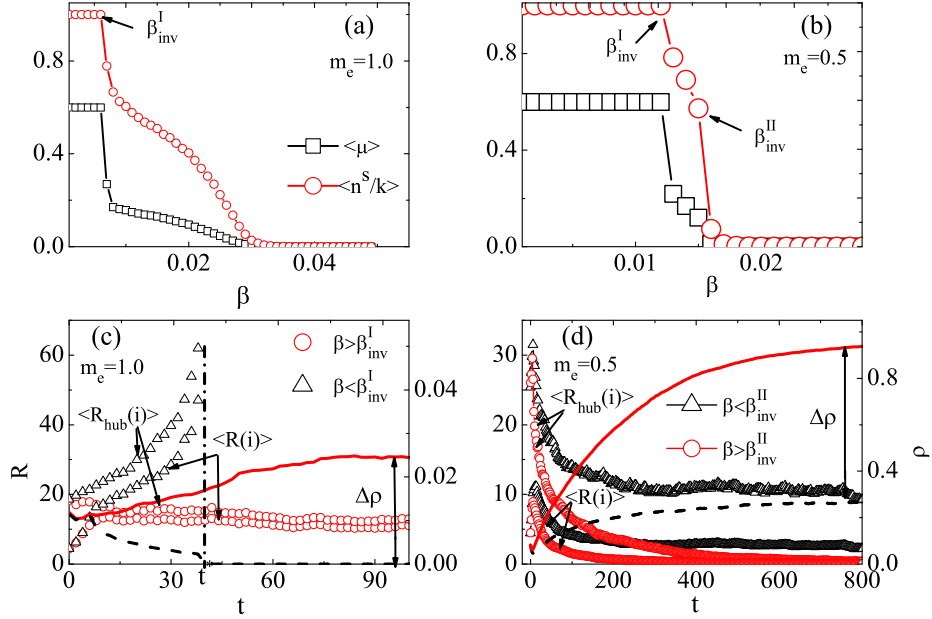}
\caption{(Color online) Analysis of the phase transition with edge
  overlap between the two layers. (a). The fraction of healthy (S)
  neighbors around infected nodes $\langle n_s/k\rangle$ (red circles)
  and average recovery rate $\langle\mu\rangle$ (black squares) as
  functions of $\beta$ in the steady state for $\gamma=2.2$ and
  $m_e=1.0$, initial infected density is $\rho(0)=0.01$. Inset is the
  case of $m_e=0.5$. (b). Time evolution of $\rho(t)$ for $\beta$ close
  to $\beta_{\rm inv}^{II}$, $\Delta\rho$ is the jump of $\rho$ in the
  steady state for $\beta$ is close below $\beta_{\rm inv}^{II}$ and
  close above $\beta_{\rm inv}^{II}$. (c). Time evolution of average
  resource of all infected nodes $\langle R(t)\rangle$, hub nodes
  $\langle R_{hub}(t)\rangle$, and infected density $\rho(t)$. $t^{*}$
  is the moment when all the neighbors of the infected nodes are in
  healthy state (there is no definition of $\langle R_{hub}(t)\rangle$
  and $\langle R(t) \rangle$). (d). The average resource of $\langle
  R(t)\rangle$ and $\langle R_{hub}(t)\rangle$ as functions of $t$.}
\label{Evolution(edgelap)}
\end{center}
\end{figure}

To clarify these results, Figs.~\ref{Evolution(edgelap)}(a) and (b) show
a plot of the average recovery rate $\langle\mu\rangle$ and the number
of susceptible neighbors around each infected individual $\langle n_s/k
\rangle$ as functions of $\beta$. Note that when the two layers overlap
completely ($m_e=1.0$), $\langle n_s/k \rangle$ and $\langle\mu\rangle$
decrease at the first threshold $\beta_{\rm inv}^{I}$ to a certain value
and then decrease continuously to zero, indicating that the infected
density in the steady state increases continuously up to 1 as $\beta$
increases. In contrast, when $m_e=0.5$ there are two abrupt jumps of
$\langle n_s/k \rangle$ and $\langle\mu\rangle$ at $\beta_{\rm inv}^{I}$
and $\beta_{\rm inv}^{II}$, respectively. Here $\langle \mu \rangle$
jumps sharply to zero at $\beta_{\rm inv}^{II}$ [see
  Fig.~\ref{Evolution(edgelap)}(b)] indicating an explosive jump in
$\rho$.

Figures~\ref{Evolution(edgelap)}(c) and \ref{Evolution(edgelap)}(d) show
the time dependence of the infected density and the resource
value. Figure~\ref{Evolution(edgelap)}(c) shows the average resource of
all infected nodes $\langle R(t)\rangle$ and the average resource of hub
nodes $\langle R_{h}(t)\rangle$ as a function of $t$ for $m_e=1.0$. Note
that when $\beta$ is immediately below $\beta_{\rm inv}^{I}$ both
$\langle R_{h}(t)\rangle$ and $\langle R(t)\rangle$ increase
continuously until $t=t^*$. When $t>t^*$ there is no definition of
resource because all infected nodes recover [see $\rho(t)$ for
  $\beta<\beta_{\rm inv}^{I}$ at $t=t^*$]. When $\beta\geq\beta_{\rm
  inv}^{I}$ the infection and recovery rates are balanced, and both
$\langle R_{h}(t)\rangle$ and $\langle R(t)\rangle$ fluctuate around a
finite value when $t\rightarrow t_{\infty}$. Thus all the infected nodes
recover with a certain probability and $\rho(t)$ also fluctuates around
a finite value when $t\rightarrow t_{\infty}$ [see $\rho(t)$ for
  $\beta>\beta_{\rm inv}^{I}$]. With an increase in $\beta$, resource
availability decreases continuously as the number infected nodes
increases until the disease spreads throughout the system and no
available resources remain [see Figs.~\ref{Evolution(edgelap)}(a) and
  \ref{Evolution(edgelap)}(c)].  This accounts for the continuous
increase in $\rho$ when $m_e=1.0$.

Figure~\ref{Evolution(edgelap)}(d) shows the time evolution of the
infected density and available resource level for $m_e=0.5$. Note that
when $\beta$ is immediately below $\beta_{\rm inv}^{II}$ at the early
stage the disease propagates within the local range of seed nodes, and
there are sufficient healthy neighbors in layer $S$ to temporarily
suppress the spread. This causes a brief increase in available resources
at the beginning of the propagation process and a slight decline in the
density of infection. Subsequently the disease rapidly spreads along the
edges in layer $C$. When edges in layer $S$ link out ($m_e=0.5$), with a
high probability that infected nodes in layer $S$ infect their neighbors,
$\langle R_{h}(t)\rangle$ and $\langle R(t)\rangle$ rapidly decline, and
$\rho(t)$ rapidly increases. Eventually infection and recovery become
balanced, and $\rho(t)$, $\langle R_h(t)\rangle$, and $\langle
R(t)\rangle$ converge to finite values. When $\beta>\beta_{\rm
  inv}^{II}$ there is also a temporary increase in both the available
resources and the density of infection. However, when the propagation
begins, unlike when $m_e=1.0$ [see Figs.~\ref{Evolution(struct)}(c) and
  \ref{Evolution(struct)}(d)] there is no balanced period at the
beginning of the process. The infection of the $S$-state nodes reduces
the resource available to a large number of infected nodes in layer $S$
and delays their recovery. This recovery delay further increases the
transmission probability in layer $C$. Thus $\langle R_h(t)\rangle$ and
$\langle R(t)\rangle$) decline sharply to zero, the density of infection
rapidly increases to 1, and cascading infection occurs.

An increase in the overlap between two layers indicates an increase in
the local social circle of an individual. When an individual's
colleagues (those frequently in contact, defined as the contact layer)
and friends (the social relations, defined as the social layer) are the
same group of people, the links in these two layers largely
overlap. When $\beta<\beta_{\rm inv}^{I}$, seed nodes initially transmit
the disease only to immediate neighbors with whom they are in frequent
contact. This high-value local effect causes infected nodes to have a
higher probability of linking with other infected nodes in layer $S$ and
lowers the level of resources available from neighbors. Thus the overlap
between two layers increases network fragility against the invasion of the disease,
and increases the probability of an epidemic breakout,
and thus lowers the epidemic threshold $\beta_{\rm inv}^{I}$.
In contrast, a lower value of overlap
rate between the two layers indicates a more global social circle,
neighbors of nodes in the social layer differ from neighbors in the
contact layer. The infected nodes in the contact layer can acquire resources
from healthy neighbors in the social layer. Thus the network is more robust
against the invasion of the disease, and there is a relatively high
epidemic threshold $\beta_{\rm inv}^{I}$. This is the reason $\beta_{\rm
  inv}^{I}$ decreases as $m_e$ increases, as shown in
Fig.~\ref{hysteresis(edgeLap)}.  When $\beta_{\rm
  inv}^{I}\leq\beta<\beta_{\rm inv}^{II}$, hub nodes promote disease
transmission, the disease breaks out in a finite range, a sufficient
number of healthy neighbors are present in layer $S$ to help infected
nodes to recover, and infection and recovery remain
balanced. Figure~\ref{Evolution(edgelap)}(d) shows that the value of
resource availability fluctuates around a finite value when
$t\rightarrow t_{\infty}$, and the density of infection converges
continuously to a finite value ($\rho\simeq0.28$). Thus in this region
the global connections in a social layer have an advantage over the
local connections [see Fig.~\ref{edgelap(capa)}(a)]. When
$\beta\geq\beta_{\rm inv}^{II}$ the disease breaks out rapidly and
globally, and the balance between infection and recovery is broken. When
$m_e<m_e^c$ (a relatively low overlap rate), the connections in layer
$S$ are more global. The infection of a small number of S-state nodes
in layer $C$ influences the recovery of a large number of I-state nodes
in layer $S$. Thus there is a delay in the recovery of infected nodes
that further increases the transmission probability, promotes the
disease spreading in layer $C$, and causes global cascading failure.
This explains the increase in $\beta_{\rm inv}^{II}$ with $m_e$ and the
explosive jump of $\rho$ [see Fig.~\ref{Evolution(edgelap)} (c)] when
$m_e<m_e^c$. In contrast, when $m_e>m_e^c$, the connections of layer $S$
is more localized, and the infection of nodes in layer $C$ delays the
recovery of the infected nodes within only a small range in layer $S$.
This small range in recovery delay does not globally increase the
effective transmission probability. Thus as the effective
transmission probability gradually increases the value of $\rho$
smoothly increases with $\beta$ [see
Figs.~\ref{Evolution(edgelap)}(a) and \ref{Evolution(edgelap)}(c)].

\section{Conclusions}
We have investigated how the level of social support affects spreading
dynamics using the susceptible-infected-susceptible model in
social-contact coupled networks. Links in the social layer represent
relationships between friends or families through which healthy nodes
allocate recovery resources to infected neighbors. Links in the contact
layer represent daily physical contacts through which the disease can
spread. Infected nodes do not have resources, and their recovery depends
on obtaining resources in layer $S$ from healthy neighbors. We assume
the recovery rate of an infected node to be a function of the resources
received from healthy neighbors. We use the DMP method to analyze the
spreading dynamics. We first examine how degree heterogeneity impacts
disease spreading. We find that degree heterogeneity enhances disease
spreading, and due to the existence of hub nodes there is a
balanced interval $\beta_{\rm inv}^{I}<\beta<\beta_{\rm inv}^{II}$ in
which the infection and recovery processes remain balanced.
The value of $\rho$
increases continuously from 0 to a finite value at the first invasion
threshold $\beta_{\rm inv}^{I}$, increases slowly in $\beta_{\rm
  inv}^{I}<\beta<\beta_{\rm inv}^{II}$, then suddenly jumps at
$\beta_{\rm inv}^{II}$.  Thus the transition of $\rho$ is hybrid.  In
addition, increasing the degree exponent $\gamma$ in the network
increases the gap between the two thresholds and the hysteresis
region. To analyze the sudden jump of $\rho$ and the hysteresis loops,
we examine the spreading process analytically using mean-field
approximation in RRNs. Through a bifurcation analysis we account for the
existence of the sudden jump of $\rho$ and the hysteresis loops. In
addition, in the RRNs the balanced interval disappears when there is a
lack of hub nodes. The first invasion threshold $\beta_{\rm inv}^{I}$ thus
disappears.

We next fix the degree heterogeneity and investigate the effect of edge
overlap between the two layers. We find that there is a critical value
$m_e^c$. When $m_e<m_e^c$ there is a second invasion threshold
$\beta_{\rm inv}^{II}$ that increases with $m_e$. The value of $\rho$
smoothly increases at $\beta_{\rm inv}^{I}$ and then suddenly jumps at
$\beta_{\rm inv}^{II}$, revealing the transition of $\rho$ to be hybrid
with the presence of hysteresis loops in this region [see
  Fig.~\ref{hysteresis(edgeLap)}]. In contrast, when $m_e>m_e^c$ the
phase transition of $\rho$ is continuous and the hysteresis loops
disappear. In addition, when $\beta<\beta_{\rm inv}^{I}$ seed nodes can
only transmit the disease locally at the early stage.  Here an increase
in global connectivity with a lower rate of overlap in the social layer
(layer $S$) increases the probability of linking to healthy neighbors
and increases the probability that infected nodes will recover. Thus the
first invasion threshold $\beta_{\rm inv}^{I}$ decreases as the overlap
rate $m_e$ increases. When $\beta>\beta_{\rm inv}^{I}$, increasing the
transmission rate increases the fraction of infected nodes, and an
increase in global connectivity in layer $S$ increases the probability
of linking to infected neighbors and lowers the recovery rate.  Thus the
second invasion threshold $\beta_{\rm inv}^{II}$ increases with $m_e$
when $m_e<m_e^c$.

Although researchers in different scientific fields have
focused on ways of constraining disease epidemics in human populations,
most scientific literature has been devoted to questions concerning the
optimum allocation of public resources or the impact of government
investment on spreading dynamics. There has been little examination of
how social supports affect spreading dynamics, and our novel model fills
this gap. In future research on the impact of social supports on
spreading dynamics we will focus on such elements as how degree
correlation and the clustering coefficient affect epidemic spreading.
Other related topics will include the effect of preference-driven resource
allocation on spreading dynamics and the interplay between disease
dynamics and resource dynamics.

\section*{Acknowledgements}
This work was supported by the National Natural
Science Foundation of China under Grants No. 11575041 and Projects
No. 61673086, No.61672238,
the Fundamental Research Funds for the Central Universities under Grant No.
ZYGX2015J153. LAB is supported by UNMdP and FONCyT Pict 0429/13

\section*{Appendix}

When $\gamma\rightarrow\infty$ the system is approximately a random
regular network (RRN). To analyze the hysteresis loop and the sudden
jump of $\rho$, we solve Eqs.~(\ref{dynamic}) and (\ref{cavityRate})
analytically for RRN using mean-field approximation. In the mean-field
approximation for a RRN, the degree of each node has the same value and
the same probability of being infected. Because we have only considered
the case $k_i^S=k_i^C$, for simplicity we denote the degree to be
$k$. Each edge in the network also has the same probability of
connecting with infected neighbors.  Thus we define $\rho(t)$ and
$\theta(t)$ such that $\rho(t)=\rho_i(t)=\rho_j(t)$ and
$\theta(t)=\theta_{j\rightarrow i}(t)=\theta_{l\rightarrow h}(t)$.
Consequently the resource that each infected node can receive from
healthy neighbors is
\begin{equation}\label{uniformR}
  R(t)=\frac{k(1-\theta(t))}{(k-1)\theta(t)+1},
\end{equation}
and the recovery rate of each node is
\begin{equation}\label{uniformMu}
  \mu(t)=\frac{\mu_r(1-\theta(t))}{(k-1)\theta(t)+1}.
\end{equation}
When we approximate $1-(1-\beta\theta(t))^k$ as $k\beta\theta(t)$ and
$1-(1-\beta\theta(t))^{(k-1)}$ as $(k-1)\beta\theta(t)$ for small
$\beta$ we obtain
\begin{equation}
\frac{d\rho(t)}{dt}=k\beta\theta(t)(1-\rho(t))
-\frac{\mu_r(1-\theta(t))}{(k-1)\theta(t)+1}\rho(t)
\label{RRNdynamic}
\end{equation}
and
\begin{equation}
\frac{d\theta(t)}{dt}=\beta(k-1)\theta(t)(1-\theta(t))
-\frac{\mu_r(1-\theta(t))}{(k-1)\theta(t)+1}\theta(t).
\label{RRNcavityRate}
\end{equation}
The steady state of the spreading process corresponds to conditions
$d\rho(t)/dt=0$ and $d\theta(t)/dt=0$. We denote $\theta(\infty)$ as $\theta$ and obtain
%
%
%
\begin{equation}
\theta(1-\theta)[\beta(k-1)-\frac{\mu_r}{(k-1)\theta+1}]=0.
\label{steadytheta}
\end{equation}
We also define $g(\theta)$ as the function of $\theta$ in the steady
state, which is
\begin{equation}
g(\theta)=\theta(1-\theta)[\beta(k-1)-\frac{\mu_r}{(k-1)\theta+1}].
\label{steadyg}
\end{equation}
Here $g(\theta)$ is tangent to the horizontal axis at
$\theta_c(\infty)$, which is the critical value in the limit
$t\rightarrow\infty$. The critical condition is
\begin{equation}
  \frac{dg(\theta)}{d\theta}|_{\theta_c}=0.
  \label{critical}
\end{equation}

\begin{figure}
\begin{center}
\includegraphics[width=0.5\linewidth]{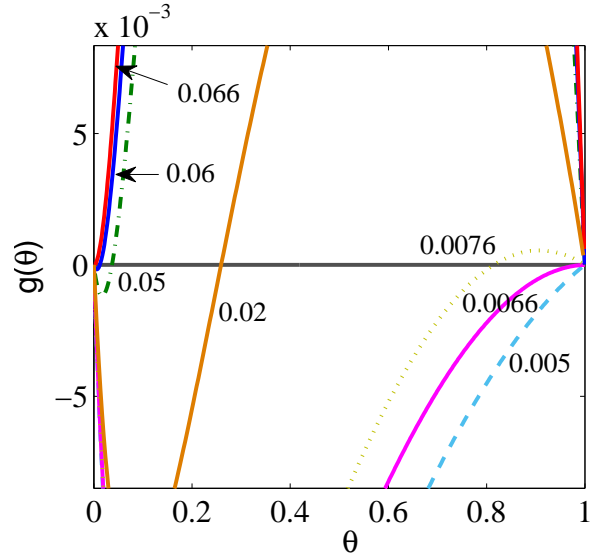}
\caption{(Color online). Illustration of graphical solution of
  Eq. (\ref{steadyg}) for RRNs with $k=10$. The red line corresponding
  to $\beta=0.066$ is tangent to the horizontal axis at $\theta=0$. The
  purple line correspionding to $\beta=0.0066$ is tangent to the
  horizontal axis at $\theta=1$}
\label{bifurcation}
\end{center}
\end{figure}

Solving Eq.~(\ref{critical}) we also obtain the critical transmission
rate.  From Eq.~(\ref{steadytheta}) we see that $\theta=1$ and
$\theta=0$ are two trivial solutions.  Figure~\ref{bifurcation} shows
that the number of solutions for Eq.~(\ref{steadyg}) is dependent on
$\beta$ and there is a critical value of $\beta$ at which three roots of
Eq.~(\ref{steadyg}) emerge, implying that a cusp bifurcation occurs. A
bifurcation analysis \cite{strogatz1994nonlinear} of Eq.~(\ref{steadyg})
indicates that the physically meaningful stable solution of $\theta$
will suddenly increase, and there is an alternate outcome---explosive
growth in $\rho$. Whether the unstable state stabilizes to an outbreak
state ($\theta>0$,$\rho>0$) or an extinct state ($\theta=0$,$\rho=0$)
depends on the initial infection density $\rho(0)$, thus a hysteresis
loop emerges. To distinguish the two thresholds of the hysteresis loop,
we denote $\beta_{\rm per}$ as the persistence threshold corresponding
to the nontrivial solution $\theta_c>0$ of Eq.~(\ref{steadyg}) at which
the disease initially has a large $\rho(0)$ value. Here $\beta_{\rm
  inv}$ is the invasion threshold corresponding to the nontrivial
solution $\theta_c=0$ of Eq.~(\ref{steadyg}) at which the disease
initially has a small $\rho(0)$ value.  The interval $[\beta_{\rm
    per},\beta_{\rm inv})$ is the hysteresis region.

Figure~\ref{bifurcation} shows an example illustrating the relationship
between $\rho$ and $\beta$ when $k=10$. Note that $g(\theta)$ is tangent
to the horizontal axis at $\theta=1.0$ when $\beta_{\rm
  per}\simeq0.0066$ and at $\theta=0.0$ when $\beta_{\rm
  inv}\simeq0.066$ respectively. When $0.0066\leq\beta<0.066$ three
roots of Eq.~(\ref{steadyg}) emerge, indicating that a saddle-node
bifurcation occurs and the physically meaningful stable solution of
$\theta$ increases suddenly to 1.  If the disease initially has a
relatively small infection density, e.g., $\rho(0)=0.01$, the system
converges to the stable state $\rho=0$, which corresponds to $\theta=0$.
On the other hand if the disease initially has a relatively large
infection density, e.g., $\rho(0)=0.9$, the system converges to the
stable state $\rho=1$, which corresponds to $\theta=1$.  When
$\beta\geq0.0066$ and $\beta<0.0066$, $\rho(0)$ has no effect on the
stable state of the system.  Thus $\beta=0.0066$ is the persistence
threshold $\beta_{\rm per}$ and $\beta=0.066$ is the invasion threshold
$\beta_{\rm inv}$.

\begin{figure}
\begin{center}
\includegraphics[width=0.5\linewidth]{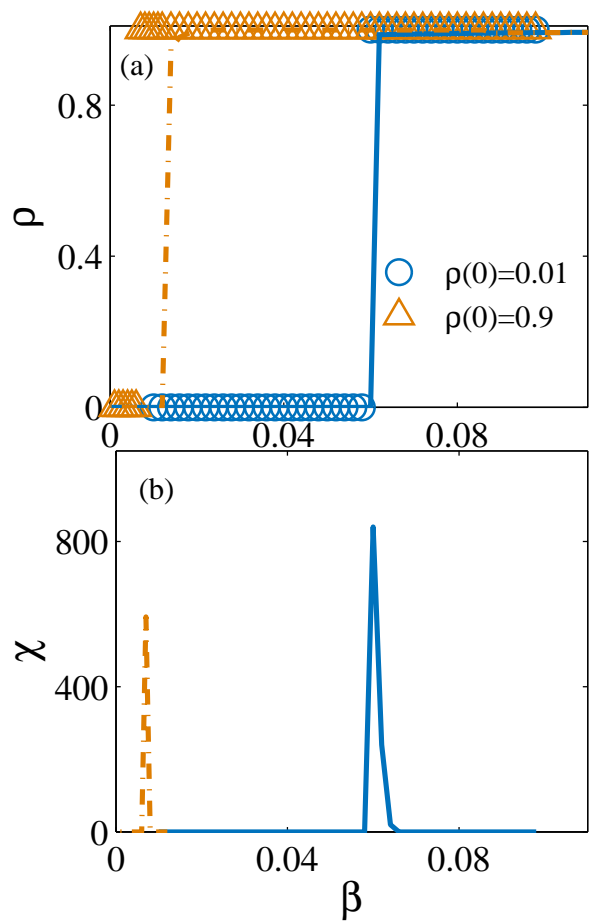}
\caption{(Color online). Phase transition of infected
    density in steady state and susceptibility measure $\chi$ on random
  regular networks. (a) Infected density $\rho$ vs $\beta$ for
  $\rho(0)=0.01$ (blue circles) and $\rho(0)=0.9$ (orange
  triangles). (b) Susceptibility measure $\chi$ vs $\beta$ for
  $\rho(0)=0.01$ (blue line) and $\rho(0)=0.9$ (orange dash
  line). Network size $N=10000$ and degree $k=10$. The
    analytical results are obtained from the mean-field approximation
    [(Eqs. (\ref{RRNdynamic}) and (\ref{RRNcavityRate}))]. }
\label{RRN(capa)}
\end{center}
\end{figure}

Figure~\ref{RRN(capa)}(a) shows the numerical and simulation results in
RRNs with a degree $k=10$.  In RRNs the first invasion threshold
$\beta_{\rm inv}^I$ disappears and the transition of $\rho$ is
discontinuous, i.e., not hybrid, due to the lack of the hub nodes.  In
addition the hysteresis loops exist in the transition of $\rho$.  The
orange dashed line and the blue line correspond to the theoretical
results for $\rho(0)=0.01$ and $\rho(0)=0.9$, respectively, obtained
from Eqs.~(\ref{RRNdynamic}) and (\ref{RRNcavityRate}).
Figure~\ref{RRN(capa)}(b) shows the susceptibility measurement $\chi$ vs
$\beta$ for $\rho(0)=0.01$ and $\rho(0)=0.9$.  From these results we
find that the theoretical results obtained from the mean-field
approximation agree with the simulation results in RRNs.

\section*{References}

\begin{thebibliography}{100}
\bibitem{meyers2005network}
Meyers L A, Pourbohloul B, Newman M E, Skowronski D M and Brunham R C 2005
J. Theor. Biol. \textbf{232} 71.

\bibitem{yuen1998clinical}
Yuen K, Chan P, Peiris M, Tsang D, Que T, Shortridge K, Cheung P, et al.,
1998 The Lancet \textbf{351} 467.

\bibitem{de2006fatal}
de Jong M D, Simmons C P, Thanh T T, et al., 2006 Nat. Med. \textbf{12} 1203.

\bibitem{team2014ebola}
Team W E R 2014 N. Engl. J. Med. \textbf{2014} 1481.

\bibitem{gallup2001economic}
Gallup J L and Sachs J D 2001 AM J. Trop. Med. Hyg. \textbf{64} 85.

\bibitem{kirigia2009economic}
Kirigia J M, Sambo L G, Yokouide A, Soumbey-Alley E, Muthuri L K and Kirigia D G 2009 BMC International Health and Human
Rights \textbf{9} 8.

\bibitem{stinnett1996mathematical}
Stinnett A A and Paltiel A D, 1996 Journal of Health Economics \textbf{15} 641.

\bibitem{wang1999role}
Wang L Y, Haddix A C, Teutsch S M and Caldwell B 1999 AM. J. Manag. Care \textbf{5} 445.

\bibitem{zaric2002dynamic}
Zaric G S and Brandeau M L 2002 Math. Med. Biol. \textbf{19} 235.

\bibitem{brandeau2005allocating}
Brandeau M L, 2005 \emph{in Operations Research and Health Care} (Springer).

\bibitem{brandeau2003resource}
Brandeau M L, Zaric G S and Richter A 2003 Journal of Health Economics \textbf{22} 575.

\bibitem{bottcher2015disease}
B{\"o}ttcher L, Woolley-Meza O, Ara{\'u}jo N A, Herrmann H J and Helbing D 2015 Sci. Rep. \textbf{5} 16571

\bibitem{araujo2010explosive}
Ara{\'u}jo N A and Herrmann H J 2010 Phys. Rev. Lett. \textbf{105} 035701.

\bibitem{nagler2012continuous}
Nagler J, Tiessen T and Gutch H W 2012 Phys. Rev. X \textbf{2} 031009.

\bibitem{d2015anomalous}
DSouza R M and Nagler J 2015 Nat. Phys. \textbf{11} 531.

\bibitem{chen2016crossover}
Chen X L, Yang C, Zhong L F and Tang M 2016 Chaos \textbf{26} 083114.

\bibitem{boccaletti2016explosive}
Boccaletti S, Almendral J, Guan S et. al. 2016 Phys. Rep. \textbf{660} 1.

\bibitem{chen2016critical}
Chen X L, Zhou T, Feng L, Yang C, Wang M M, Fan X M and Hu Y Q, 2016 arXiv:1611.00212.

\bibitem{bottcher2016connectivity}
B{\"o}ttcher L, Woolley-Meza O, Goles E, Helbing D and Herrmann H 2016 Phys. Rev. E \textbf{93} 042315.

\bibitem{seeman1996social}
Seeman T E 1996 Ann. Epidemiol. \textbf{6} 442.

\bibitem{schulz2008physical}
Schulz R and Sherwood P R 2008 Journal of Social Work Education \textbf{44} 105.

\bibitem{drummond2015methods}
Drummond M F, Sculpher M J, Claxton K, Stoddart G L and Torrance G W 2015 \emph{Methods for the economic evaluation of health care
programmes} (Oxford university press).

\bibitem{cohen1985social}
Cohen S E and Syme S, 1985 \emph{Social support and health} (Academic Press).

\bibitem{thoits1995stress}
Thoits P A 1995 Journal of Health and Social Behavior \textbf{2} 5379.

\bibitem{gomez2013diffusion}
Gomez S, Diaz-Guilera A, Gomez-Gardenes J, Perez-Vicente C J, Moreno Y and Arenas A 2013 Phys. Rev. Lett. \textbf{110}, 028701.

\bibitem{granell2013dynamical}
Granell C, Gomez S and Arenas A 2013 Phys. Rev. Lett. \textbf{111} 128701.

\bibitem{bianconi2016percolation}
Bianconi G and Radicchi F 2016 arXiv:1610.08708.

\bibitem{de2016physics}
De Domenico M, Granell C, Porter M A and Arenas A 2016  Nat. Phys. \textbf{12} 901906.

\bibitem{karrer2010message}
Karrer B and Newman M E 2010 Phys. Rev. E \textbf{82} 016101.

\bibitem{shrestha2014message}
Shrestha M and Moore C 2014 Phys. Rev. E \textbf{89} 022805.

\bibitem{shrestha2015message}
Shrestha  M, Scarpino S V and Moore C 2015 Phys. Rev. E \textbf{92} 022821.

\bibitem{wang2016unification}
Wang W, Tang M, Stanley H E and Braunstein  L A 2017 Rep. Prog. Phys. \textbf{80} 036603.

\bibitem{guimera2005worldwide}
Guimera R, Mossa S, Turtschi A and Amaral L N 2005 Proc. Natl. Acad. Sci. USA \textbf{102} 7794.

\bibitem{bullmore2012economy}
 Bullmore E and Sporns O 2012 Nat. Rev. Neurosci.\textbf{13} 336.

\bibitem{valdez2016failure}
 Valdez L D, Di Muro M A and Braunstein L A 2016 J. Stat. Mech-Theory E \textbf{2016} 093402.

 \bibitem{PastorSatorras2001}
Pastor-Satorras  R and Vespignani A 2001 Phys. Rev. Lett. \textbf{86} 3200.

\bibitem{pastor2015epidemic}
Pastor-Satorras R, Castellano C, Van Mieghem P and Vespignani A 2015 Rev. Mod. Phys. \textbf{87} 925.

\bibitem{gomez2010discrete}
Gomez S, Arenas A, Borge-Holthoefer J, Meloni S and Moreno Y 2010 Europhys. Lett. \textbf{89} 38009.

\bibitem{krzakala2013spectral}
 Krzakala  F, Moore C, Mossel E, Neeman J, Sly A, Zdeborova L and Zhang P 2013 Proc. Natl. Acad. Sci. USA \textbf
{110} 20935.

\bibitem{viswanath2009evolution}
Viswanath B, Mislove A, Cha M and Gummadi K P 2009 Proceedings of the 2nd ACM workshop on Online social networks pp: 37-42.

\bibitem{adamic2000power}
Adamic L A and Huberman B A 2000 Science \textbf{287} 2115.

\bibitem{catanzaro2005generation}
Catanzaro M, Bogu{\~n}{\'a} M and Pastor-Satorras R 2005 Phys. Rev. E \textbf{71}, 027103.

\bibitem{boguna2004cut}
Boguna  M, Pastor-Satorras R and Vespignani A 2004
Euro. Phys. J. B \textbf{38} 205.

\bibitem{pastor2001epidemicprl}
Pastor-Satorras R and Vespignani A 2001 Phys. Rev. Lett. \textbf{86} 3200.

\bibitem{ferreira2012epidemic}
Ferreira S C, Castellano C and Pastor-Satorras R 2012 Phys. Rev. E \textbf{86} 041125.

\bibitem{shu2016recovery}
Shu P, Wang W, Tang M, Zhao P and Zhang Y C 2016 Chaos \textbf{26} 063108.

\bibitem{nagler2011impact}
Nagler J, Levina A and Timme M 2011 Nat. Phys. \textbf{7}, 265.

\bibitem{newman1999monte}
Newman M and Barkema G 1999 \emph{Monte Carlo Methods in Statistical Physics} (Oxford University Press).

\bibitem{cellai2013percolation}
Cellai D, Lopez E, Zhou J, Gleeson J P and Bianconi G 2013 Phys. Rev. E \textbf{88} 052811.

\bibitem{strogatz1994nonlinear}
Strogatz S 1994 Nonlinear dynamics and chaos: With applications (Addison-Wesley, Reading, MA).

\end{thebibliography}
\end{document}